\documentstyle[12pt]{article}

\newcommand\beq{\begin{equation}}
\newcommand\eeq{\end{equation}}
\def\beqa{\begin{eqnarray}}
\def\eeqa{\end{eqnarray}}
\def\bega{\begin{array}}
\def\enda{\end{array}}
\def\non{{\nonumber}}

\def\|{\'\i}

\begin{document}

\title{The Negative Dimensional Oscillator at Finite Temperature}
\author{Silvio J. Rabello\thanks{e-mail:IFT10034@UFRJ.BITNET},
Arvind N. Vaidya\\
and\\ Luiz Claudio M. de Albuquerque\\
\\{\it Instituto de F\|sica}\\
{\it Universidade Federal do Rio de Janeiro}\\
{\it Rio de Janeiro  RJ}\\
{\it Caixa Postal 68.528-CEP 21945-970}\\
{\it Brasil}}
\date{}
\maketitle
\vspace{-11cm}
\hfill IF-UFRJ-November 93
\vspace{11cm}
\begin{abstract}

{\sl We study the thermal behavior of the negative dimensional harmonic
oscillator of Dunne and Halliday that at zero temperature, due to a
hidden BRST symmetry of the classical harmonic oscillator, is shown to be
equivalent to the Grassmann oscillator of Finkelstein and Villasante.
At finite temperature we verify that although being described by Grassmann
numbers the thermal behavior of the negative dimensional oscillator is
quite different from
a Fermi system.}

\end{abstract}
\newpage
%%%%%%%%%%%%%%%%%%%%%%%%%%%%%%%%%%%%%%%%%%%%%%%%%%%%%%%%%%%%%%%%%%%%%%%%%%%

It is generally accepted that Grassmann variables correspond to
negative dimensional degrees of freedom. For example, in the
Becchi-Rouet-Stora-Tyutin (BRST)
quantization of constrained dynamical systems \cite{HeTe} the ghosts cancel
the unphysical degrees of freedom  and in the Parisi-Sourlas \cite{PaSo}
approach to scalar fields with random sources we have a dimensional
reduction by the use of Grassmann variables, what was later extended by
McClain et al \cite{McNi} to gauge
and Fermi fields. In their study of  Feynman integrals continued to
negative dimensions, Dunne and Halliday \cite{DuHa} pointed that the
Grassmann Oscillator (GO) of Finkelstein and Villasante \cite{FiVi} share
the spectrum and
degeneracy of a harmonic oscillator (HO) continued to
negative values of the dimension. Also Dunne \cite{Du} has shown that it is
possible to define negative dimensional classical groups  by their action
on Grassmann representation spaces. More recently Delbourgo et al
\cite{DeJoWh}
studied both the anharmonic Grassmann oscillator and the GO,
and used the GO as a guide to  construct an improved Dirac equation
\cite{DeJo}
in a space with both commuting and anticommuting coordinates, obtaining
a discrete and finite mass spectrum that could explain the families of
quarks and leptons origin.

In this letter we apply the concept of classical path integral of Gozzi
\cite{Go}
and Gozzi et al \cite{GoRe} to the HO, and show that at zero temperature the
cancellation of degrees of freedom between the HO and GO is due to a hidden
BRST symmetry of the classical HO. At finite temperature we have verified
that this BRST symmetry induces a ghost-like thermal behavior for the
Grassmann degrees of freedom.

In ref.\cite{FiVi} Finkelstein and Villasante used the Schwinger action
principle as a guide to introduce a Grassmann generalization of ordinary
quantum
mechanics, where the degrees of freedom are elements of a Grassmann algebra.
Their approach was different from the previous work on Grassmann and
supersymmetric quantum mechanics in the sense that their theory is
second order in time derivatives instead of the first order models based
on field theoretical fermions. As an example they studied the generalization
of the harmonic oscillator by introducing an analogous anticommuting system
described by N pairs of canonically conjugate operators $\hat q_a $
and $\hat p_a $ (a=1,...N) with the Halmiltonian operator and
anticommutation relations given by:
\beq
\label{H}
\hat H=-{1\over 2}\hat p_aC_{ab}^{-1}\hat p_b+{{\omega^2}\over2}
\hat q_a{C_{ab}}\hat q_b.
\eeq
and $(\rm\hbar=1)$
\beq
\label{A-comm}
[\hat q_a,\hat p_b]_{+}=-i\delta_{ab}\,,\qquad[\hat q_a,\hat q_b]_{+}=0\,,
\qquad[\hat p_a,\hat p_b]_{+}=0
\eeq
Here there is  a summation over repeated indices from 1 to N,
C is a hermitian antisymmetric matrix chosen to make ${\rm\hat H}$ self
conjugate, and as an antisymmetric matrix  has an inverse only for even
dimensions, we are restricted to even values of N. For an isotropic
oscillator we can always put C in a block-diagonal form, with block
elements given by the Pauli matrix $\sigma_2$:

\beq
\label{Cab}
\sigma_2=\pmatrix{0 & -i\cr i&0\cr}
\eeq
Due to the fact that conjugation reverses the order of the
operators we have that  $\hat q_a$ and $\hat p_a$ cannot be both self
conjugate and still satisfy (2), so we choose $\hat q_a=\hat {q_a}
^\dagger$ and $\hat p_a=-\hat{p_a}^\dagger$.

The Heisenberg equations of motion are given by {\it i} times the
commutator of $\hat H$ with $\hat{q}_a$ and $\hat{p}_a$ as in ordinary
quantum mechanics:

\beq
\label{Heis}
{{d\hat q_a}\over{dt}}=i[\hat H,\hat q_a]_{-}=C^{-1}_{ab}\hat p_b\,,\qquad
\qquad {{d\hat p_a}\over{dt}}=i[\hat H,p_a]_{-}=-\omega^2 C_{ab}
\hat q_b
\eeq
which have oscillatory solutions of frequency $\omega$ as for
the bosonic HO.

To see how the degrees of freedom of the GO and the HO cancel each other
we study  a system where both of them  are present, namely the path integral
formulation for the classical mechanics of the HO.
This formulation of classical mechanics was introduced by Gozzi \cite{Go}
and later developed by Gozzi et al \cite{GoRe}, as a functional formulation
of classical mechanics. Their starting point was as in quantum field theory
to write the generating functional as a functional integral over
all trajectories but with weight `one' for the classical paths and
weight `zero' to all the others. A representation of this generating
functional for a classical system with  n degrees of freedom ${x_i(t)}$
and action S[x], is given by:
\beq
\label{Zj1}
Z_{Class.}[{\bf J}]=\int[d{\bf x}]\;\delta[{\bf x}-{\bf x}_{class.}]
\;e^{\int_0^T dt {\bf J}\cdot{\bf x}}
\eeq

Where the functional delta function ensures that only the paths
${\bf x}_{class.}(t)$ satisfying ${{\delta S}\over {\delta{\bf x}}}=0$ will
contribute to the correlation functions obtained by functional
differentiation with respect to the external current {\bf J}, that is
turned on in the interval $0\leq t\leq T$. An equivalent description is:

\beq
\label{Zj2}
Z_{Class.}[{\bf J}]=\int[d{\bf x}]\;\delta[
{{\delta S}\over {\delta{\bf x}}}]\;\parallel
det\biggl({{\delta^2 S}\over{\delta x_i (t)\delta x_j(t')}}\biggr) \parallel
 \;e^{\int_0^T dt {\bf J}\cdot{\bf x}}
\eeq
Using a functional Fourier transform for the delta function
and the Faddeev-Popov trick to handle the determinant we get:

\beq
\label{Zj3}
Z_{Class.}[{\bf J}]=\int[d{\bf x}][d\lambda][d\bar c][dc]\; e^{iS_{eff}
+ \int_0^T dt {\bf J}\cdot{\bf x}}
\eeq
where
\beq
\label{Seff}
S_{eff}=\int_0^T dt\;\{\lambda_a {{\delta S}\over {\delta x_a}}
-i{\bar c_a}{{\delta^2 S}\over{\delta x^2}}c_a\}\,,
\eeq
$\lambda_a(t)$ being a real c-number field and the ghosts
$c_a(t)$ and $\bar c_a(t)$ are real Grassmann fields.

What we have here is a very interesting situation in which the above
generating functional seems to describe a quantum mechanical problem for
the extended configuration space of the variables ${ x_a}$, $\lambda_a$,
$c_a$ and $\bar c_a$ with the above $\rm S_{eff}$, although in the $x_a$
subspace we have a classical system with
${{\delta S}\over {\delta{\bf x}}}=0$. The truth is that the resulting
`quantum' system is in fact a topological field theory with zero
degrees of freedom. To understand this point notice that $\rm S_{eff}$ has
a rigid BRST symmetry:

\beq
\label{BRST}
\delta x_a=\epsilon c_a,\qquad\delta c_a=0,\qquad\delta\bar c_a=-i
\epsilon\lambda_a,\qquad\delta\lambda_a=0
\eeq
and also an anti-BRST one

\beq
\label{a-BRST}
\bar\delta x_a=-\bar\epsilon\bar c_a,\qquad\bar\delta c_a=-i\bar\epsilon
\lambda,\qquad\bar\delta\bar c_a=0,\qquad\bar\delta\lambda_a=0
\eeq
where $\bar\epsilon$ and $\epsilon$ are Grassmann parameters and
obviously $\delta^2=\bar\delta^2=0$. The above  invariances (\ref{BRST}) and
(\ref{a-BRST})
are generated respectively  by the operators $\bf s$ and $\bar{\bf s}$,
defined by:

\beq
\label{Gfix}
\delta\Phi=\epsilon{\bf s}\Phi\qquad,\qquad\bar\delta\Phi=\bar\epsilon
\bar{\bf s}\Phi
\eeq
where $\Phi$ is any functional of the extended configuration space
variables.

Using the above results it is easy to see that $\rm S_{eff}$ can be written
as a BRST variation:

\beq
\label{Div}
S_{eff}=i\int_0^T dt\;{\bf s}\,\biggl(\bar c_a{{\delta S}\over
{\delta x_a}}\biggr)
\eeq

We can now verify that our generating functional  describes a topological
theory. Recalling the BRST quantization of constrained systems \cite{HeTe},
we see
that $\rm S_{eff}$ is just a gauge fixing term for a theory of topological
action $\rm S_{Top}[{\bf x}]=0$ and invariant under the local transformation
$x_a(t)\rightarrow x_a(t)+\delta x_a(t)$ with arbitrary
$\delta x_a(t)$. In other words the system defined in the extended
configuration space ($x_a$, $\lambda_a$, $ c_a$, $\bar c_a$ )
is insensitive to arbitrary deformations of the path $x_a(t)$ and it
is in that sense that we call it topological.

Let us now consider the classical path integral for a n-dimensional harmonic
oscillator of unit mass, with action:

\beq
\label{Ho}
S=\int_0^Tdt\;{1\over 2}({\bf \dot x}^2-\omega^2{\bf x}^2)
\eeq
For this system the $\rm S_{eff}$ reads:

\beq
\label{HoGo}
S_{eff}=-\int_0^T dt\;\{\lambda_a ({{d^2}\over{dt^2}}+\omega^2)x_a
-i{\bar c_a}({{d^2}\over{dt^2}}+\omega^2)c_a\}
\eeq

\noindent As we can see our $\rm S_{eff}$ describes a 2n-dimensional bosonic
oscillator in the variables $\lambda_a(t)$ and $ x_a(t)$ and a
2n-dimensional fermionic oscillator in the ghost variables $ c_a(t)$ and
$\bar c_a(t)$. To see how this ghost oscillator is related to the
Grassmann oscillator we identify

\beq
\label{ReN}
\bar c_a=q_{2a-1}\,,\qquad c_a=q_{2a}\,,\qquad a=1,...n
\eeq
so that if we set 2n=N we get the Grassman oscillator action, up to
a total time derivative.

Our conclusion is that the GO can be thought as the ghost part of the BRST
quantization of a topological action $\rm S_{Top}[{\bf x}]=0$ with the gauge
condition that the $ x_a(t)$ obey the harmonic oscillator classical equations
of motion. As this topological action has zero degrees of freedom, it is
legitimate to think of the GO as a negative dimensional system. This
cancellation of degrees of freedom works as long as the HO+GO system
displays the BRST and anti-BRST invariances (\ref{BRST}) and (\ref{a-BRST}).
In ref. \cite{DuHa} a similar graded symmetry in the ($x$,$q$) sector
was pointed as the responsible for the positivity
of the HO+GO quantum mechanical spectrum, in ref. \cite{PaSo} a
supersymmetry was found to be responsible for the dimensional reduction of a
scalar field theory in the
presence of random external sources, and in ref. \cite {McNi} this
dimensional reduction mechanism was extended to gauge and Fermi fields.

The partition function for the negative dimensional HO can be obtained by
setting $\rm N\rightarrow-N$ in the partition function of a N-dimesional HO,
in equilibrium with a heat bath at inverse temperature $\beta$
$(\beta=1/kT)$,  \cite{La}:

\beqa
\label{Part}
Z^{HO}(\beta;-N)&=&\biggl(2\sinh(\beta{\omega\over 2})\biggr)^{N} \non \\
&=&\sum_{n=0}^{N} {N\choose n}(-1)^{n} e^{-\beta\omega(n-{N\over 2})}
\eeqa
where we can identify the energy spectrum
$E_{n}$ and degeneracy $g_{n}$ of the Grassmann oscillator [4,5]:
\beqa
\label{spectrum}
E_n&=&\omega(n-{N\over 2})\qquad n=0,...N \\
g _n&=&{N\choose n}
\eeqa

The $(-1)^{n}$ factor comes from the normalization of the GO energy
eigenstates
$\vert n,i\rangle$, i=1,...$g_n$:
\beq
\label{norm}
\langle n,i\vert m,j\rangle=(-1)^{n}
\delta_{nm}\delta_{ij}
\eeq

In terms of the above eigenstates  and the `position' eigenstates $\vert
q\rangle$ ,
the partition function can be written as:

\beqa
\label{Part2}
Z^{HO}(\beta;-N)&=&\sum_{n=0}^{N}\sum_{i=1}^{g_n}\langle n,i\vert
e^{-\beta\hat H }\vert n,i\rangle \non \\
&=&\sum_{n=0}^{N}\sum_{i=1}^{g_n} \int d^Nq\int d^Nq'\;
\langle n,i\vert q\rangle\langle q\vert
e^{-\beta\hat H } \vert q'\rangle \langle q'\vert n,i\rangle \non \\
&=&\sum_{n=0}^{N}\sum_{i=1}^{g_n} \int d^Nq\int d^Nq'\;
(-1)^{n}\langle q'\vert n,i\rangle\langle n,i\vert q\rangle\langle q\vert
e^{-\beta\hat H } \vert q'\rangle  \non \\
&=&\int d^Nq\;\langle q\vert e^{-\beta\hat H }\vert q\rangle
\eeqa
the $(-1)^{n}$ factor reappeared in the sum because the GO wave function
\cite{FiVi} ($\bar{q}_{\alpha}=q_{\beta}C_{\beta\alpha}$),
\beq
\label{wavfunc}
\langle q\vert n,i\rangle ={(-1)^{n}\over{2^{n\over 2}}}e^{{1\over2}
\omega\bar{q}q}{\partial\over{\partial\bar{q}_{\alpha_1}}}...
{\partial\over{\partial\bar{q}_{\alpha_n}}}e^{-\omega\bar{q}q}
\eeq
has Grassmann parity n mod 2.

{}From the above considerations we are led to the conclusion that the
partition function can be  obtained by setting t=$\rm-i\beta$ in
the propagator for the GO and integrating on q (in the Berezin
sense) with the boundary condition:

\beq
\label{bc}
q_a(\beta)=q_a(0)
\eeq
We see that although the negative dimensions have an interpretation in
terms of elements of a Grassmann algebra, their behavior at finite
temperature is quite different from a fermionic system that must be
anti-periodic in $\beta$ \cite{Ri}. It is easy to understand this ghost-like
behavior of negative dimensions by noticing that if the positive and
negative dimensional degrees  of freedom
satisfied different boundary conditions, there would be  no way
to preserve  both the BRST and anti-BRST rigid symmetries
, since once we set periodic boundary conditions for the bosonic fields
the relations (\ref{BRST}) and (\ref{a-BRST}) will fix the boundary
conditions of the Grassmann fields as periodic. In the case of
supersymmetric theories the different boundary conditions of bosons and
fermions leads to a spontaneous supersymmetry
breaking at finite temperature \cite{Das}.

To see how the cancellation of degrees of freedom takes place  we have that
the product of the partition functions for both the positive and negative
dimensional HO gives the identity,  so we are dealing with a
topological field theory over the circle $\rm S^1$ with length $\beta$,
and the observables obtained from $\rm Z(\beta)$ will be independent
of the `metric' factor $\beta$.

We have verified that the connection of the Grassmann oscillator to
the harmonic oscillator continued to negative values of the dimension
is due to a rigid BRST symmetry, that allows one to see the HO+GO system
as a topological theory, and also that in order to describe a
negative dimensional HO at finite temperature by the use of Grassmann
degrees of freedom we must  preserve the imaginary time periodic
boundary condition of the positive dimensional HO, expressing in
that way the ghost-like character of negative dimensions.

\newpage
{\bf Acknowledgments}
\vskip 0.5 cm

The authors are grateful to Dr Carlos Farina de Souza for reading
the manuscript and for many stimulating discussions. This work was partially
supported by CNPq (S.J.Rabello and A.N. Vaidya) and CAPES (L.C.M. de
Albuquerque).

\end{document}